\documentclass[10pt]{iopart}
\usepackage[dvips]{graphicx}
\begin{document}
\title{Ferromagnetic Kondo-Lattice Model}
\author{W.~Nolting, W.~M{\"u}ller and C.~Santos}
\address{Lehrstuhl Festk{\"o}rpertheorie,
  Institut f{\"u}r Physik, Humboldt-Universit{\"a}t zu
  Berlin, Invalidenstra{\ss}e 110, 10115 Berlin, Germany}

\begin{abstract}
We present a many-body approach to the electronic and magnetic
properties of the (multiband) Kondo-lattice model with ferromagnetic interband
exchange. The coupling between itinerant conduction electrons and
localized magnetic moments leads, on the one hand, to a distinct temperature-dependence of the electronic
quasiparticle spectrum and, on the other hand, to magnetic properties, as e.~g.~the Curie
temperature $T_{\rm{C}}$ or the magnon dispersion, which are strongly influenced
by the band electron selfenergy and therewith in particular by the
carrier density. We present results for the single-band Kondo-lattice
model in terms of quasiparticle densities
of states and quasiparticle band structures and demonstrate the
density-dependence of the self-consistently derived Curie temperature. The
transition from weak-coupling (RKKY) to strong-coupling (double exchange)
behaviour is worked out.

The multiband model is combined with a tight-binding-LMTO
bandstructure calculation to describe real magnetic materials. As an
example we present results for the archetypal ferromagnetic local-moment
systems EuO and EuS. The
proposed method avoids the double counting of relevant interactions and
takes into account the correct symmetry of atomic orbitals.
\end{abstract}

\section{Introduction}
The Kondo-lattice model (KLM), being also denoted as \textit{s-f} or \textit{s-d model} or,
in its strong-coupling regime, as \textit{double exchange model}, is
today surely one of the most prominent models
in solid state theory, and that mainly because of its great variety of
important applications to rather hot topics in the wide field of
collective magnetism. It refers to magnetic materials which get their
magnetic properties from a system of localized magnetic moments being
indirectly coupled via an interband exchange $J$ to itinerant
conduction electrons. Many characteristic properties of such materials
can be traced back to this interband exchange.

There are a lot of important applications of the KLM, first of all to the classical systems of
magnetic semiconductors such as EuO and EuS \cite{Wachter79} and magnetic
metals such as Gd \cite{DonDoNo98} ($J>0$, weak to intermediate coupling). All these materials have strictly
localized moments because of the half-filled $4f$ shell of the rare
earth ion (Eu$^{2+}$, Gd$^{3+}$). The striking temperature-dependence of
the unoccupied (!) conduction bandstates in the EuX \cite{Wachter79} manifesting
itself, e.~g., in the well-known 
\textit{red shift effect} \cite{Wachter79}, and the effective moment coupling in Gd via a
conduction electron spin polarization (RKKY) \cite{DonDoNo98} can be understood only
by the existence of the interband exchange $J$.

Other candidates are the intensively investigated diluted magnetic semiconductors such as
Ga$_{1-x}$Mn$_{x}$As with probably negative $J$ (strong coupling). The Mn$^{2+}$ ion creates simultaneously an
$S=\frac{5}{2}$ spin and an itinerant hole in the GaAs valence band
\cite{OhnoScience}. The latter takes care for an indirect coupling between the randomly
distributed magnetic moments that leads already for very small $x$ to a
ferromagnetic order with a maximum $T_{\rm{C}}$ of $110$~K for $x=0.053$
\cite{MatsukaraOhno}.

Another hot topic because of great potential for
engineering applications are the colossal
magnetoresistance (CMR) materials \cite{Ramirez97} such as
\mbox{La$_{1-x}$(Ca,Sr)$_{x}$MnO$_{3}$} which show up a very rich magnetic
phase diagram. For $0.2\leq x \leq 0.4$ the originally antiferromagnetic
and insulating parent compound LaMnO$_{3}$ changes into a ferromagnetic
metal. This is ascribed to a homogeneous valence mixing \mbox{(Mn$_{1-x}^{3+}$
Mn$_{x}^{4+}$)}. The three $3d_{t_{2g}}$ electrons in Mn$^{3+}$ are
considered more or less localized forming an $S=\frac{3}{2}$ moment while the additional
$3d_{e_{g}}$ electron is itinerant. Again, the local moment-itinerant
electron exchange is made responsible for many typical features of the
CMR materials ($J>0$, strong coupling).

We derive in the next section the multiband-Hamiltonian of the
Kondo-lattice model which is combined in Section 4 with an \textit{``ab
  initio''} bandstructure calculation to get the temperature-dependent
electronic structure of the prototype ferromagnets EuO and EuS. To
understand the physics of the KLM we present in Section 3 for the
nondegenerate case an analysis of the magnetic and electronic model properties.

\section{Multiband Kondo-Lattice Model}
The system under consideration consists of quasi-free electrons in
rather broad conduction bands (d bands) and localized electrons with
extremely flat dispersions (f levels). While there is no contribution of
the f electrons to the kinetic energy, the part of the d electrons reads
\begin{equation}
  \label{eq:kinEn}
  H_{d}=\sum_{ijm\bar{m}\sigma}T_{ij}^{m\bar{m}}c_{im\sigma}^{+}c_{j\bar{m}\sigma}
\end{equation}
The hopping process from site $\mathbf{R}_{i}$ to site $\mathbf{R}_{j}$
may be accompanied by an orbital change ($m\rightarrow
\bar{m}$). $T_{ij}^{m\bar{m}}$ are 
the respective hopping integrals. $c^{+}_{jm\sigma}$
($c_{jm\sigma}$) is the creation (annihilation) operator for a Wannier
electron at site $\mathbf{R}_{j}$ in the orbital m with spin $\sigma$
($\sigma=\uparrow$, $\downarrow$).

When applying the model study to real local-moment materials, intended
as our final goal, we have to require that the single-electron energies do not
only account for the kinetic energy and the influence of the lattice
potential, but also for all those interactions which are not explicitly
covered by the model Hamiltonian. That means that the hopping integrals
are to be taken from a proper \textit{``ab initio''} band calculation. We
exemplify the procedure in Sect.~4 for the special case of the
ferromagnetic europium chalcogenides. In particular we show how to avoid the well-known double-counting
problem of important interactions.

The Coulomb interaction is restricted to intraatomic terms. Furthermore,
we assume that only two subbands are involved in the scattering
process. Then the interaction part can be written as \cite{schiller01:_kondo} 
\begin{equation}
  \label{eq:multiham}
  H_{c}=H_{dd}+H_{ff}+H_{df}
\end{equation}
In an obvious manner the Coulomb interaction may be split into three
different parts depending on whether both interacting particles stem
from a conduction band $H_{dd}$, or both from a flat band
$H_{ff}$, or one from a flat band and the other from a conduction band
$H_{df}$. The first term refers to electron correlations in the
conduction bands, not explicitly taken into account by the KLM. As
mentioned, our procedure for describing real materials accounts for such
correlations by a proper renormalization of the single particle
energies.

The term $H_{ff}$ describes interactions between electrons
from the flat bands which interest us only with respect to the fact
that they form permanent magnetic moments (spins):
\begin{equation}
  \label{eq:localspin}
 \mathbf{S}_{i}=\sum_{f}\mbox{\boldmath$\sigma$\unboldmath}_{f}
\end{equation}
$\mbox{\boldmath$\sigma$\unboldmath}_{f}$ is the spin operator of an electron in subband f. If it is
necessary to consider a (super)exchange interaction between the local
spins then $H_{ff}$ is chosen to be a Heisenberg Hamiltonian, possibly
with a symmetry-breaking single-ion anisotropy $D$:
\begin{equation}
  \label{eq:filmheisenberg}
  H_{ff}=-\sum_{ij}J_{ij}\mathbf{S}_{i}\cdot\mathbf{S}_{j}-D\sum_{i}\left(S_{i}^{z}\right)^{2}
\end{equation}

The third term in (\ref{eq:multiham}) $H_{df}$ refers to the interaction
between a localized and an itinerant electron. Neglecting unimportant  
spin-independent contributions it can be written as an intraatomic exchange, i. e. a local
interaction between the conduction electron spin $\mbox{\boldmath$\sigma$\unboldmath}_{jm}$ and the local moment
spin $\mathbf{S}_{j}$ \cite{schiller01:_kondo} :
\begin{equation}
  \label{eq:spinflip}
  H_{df}=-J\sum_{jm}\mbox{\boldmath$\sigma$\unboldmath}_{jm}\cdot\mathbf{S}_{j}
\end{equation}
Here $m$ denotes the conduction electron orbital. $J$ is the exchange
coupling constant being assumed to be identical for all df pairs. Using second
quantization for the itinerant electron spin  ($n_{jm\sigma}=c_{jm\sigma}^{+}c_{jm\sigma}$),
 this interaction reads:
\begin{equation}
  \label{eq:flipflip}
  H_{df}=
  -\frac{1}{2}J\sum_{jm}\left(S^{z}_{j}\left(n_{jm\uparrow}-n_{jm\downarrow}\right)
    +S_{j}^{+}c_{jm\downarrow}^{+}c_{jm\uparrow}+S_{j}^{-}c_{jm\uparrow}^{+}c_{jm\downarrow}\right)
\end{equation}
We see that the first term describes an Ising-like interaction of the
two spin operators while 
the two others provide spin
exchange processes between localized moment and itinerant electron. In
particular the two latter terms are responsible
 for many typical
properties of the Kondo-lattice materials. Spin exchange may happen in
three different elementary processes: magnon emission by an itinerant
$\downarrow$ electron, magnon absorption by an $\uparrow$ electron and
formation of a quasiparticle (\textit{magnetic polaron}). The latter can
be understood as a propagating electron dressed by a virtual cloud of
repeatedly emitted and reabsorbed magnons corresponding to a
depolarization of the immediate localized spin neighbourhood.

The total Hamiltonian of the multiband-Kondo lattice is composed of
(\ref{eq:kinEn}), (\ref{eq:multiham}) and (\ref{eq:flipflip})
\begin{equation}
  \label{eq:totham}
  H=H_{d}+H_{ff}+H_{df}
\end{equation}
An important model parameter
is of course the effective coupling constant $\frac{J}{W}$ where $W$ is the
Bloch-bandwidth. Especially the sign of $J$ is decisive. Other parameters
are the local spin $S$, the lattice structure, and above all the band
occupation $n=\sum_{m\sigma}\langle n_{m\sigma}\rangle$. In case of a
nondegenerate band $n$ is a number in between 0 and 2.

\section {Magnetic and Electronic Model Properties}
The many-body problem of the KLM is not exactly solvable for the general
case even for the nondegenerate case, for which we can remove the
orbital dependence and which shall exclusively be tackled in this section:

\begin{equation}
  \label{eq:hamiltonian}
  H=\sum_{ij\sigma}T_{ij}c^+_{i\sigma}c_{j\sigma}-J\sum_{j}\mathbf{S}_{j}
  \cdot\mbox{\boldmath$\sigma$\unboldmath}_{j}=H_0+H_1
\end{equation}
However, there exists a non-trivial, very illustrative limiting case which is
rigorously tractable and nevertheless exhibits all the above-mentioned elementary
excitation processes
\cite{meyer01:_quant_kondo,NoDu85,NDM85,ShasMt81}. It refers 
to a single electron in an otherwise
empty conduction band coupled to a ferromagnetically saturated moment
system, e.~g.~EuO at $T=0$. In this case the $\uparrow$ spectrum is
simple, because the $\uparrow$ electron cannot exchange its spin with
\begin{figure}[htbp]
  \begin{center}
    \includegraphics[width=0.65\linewidth]{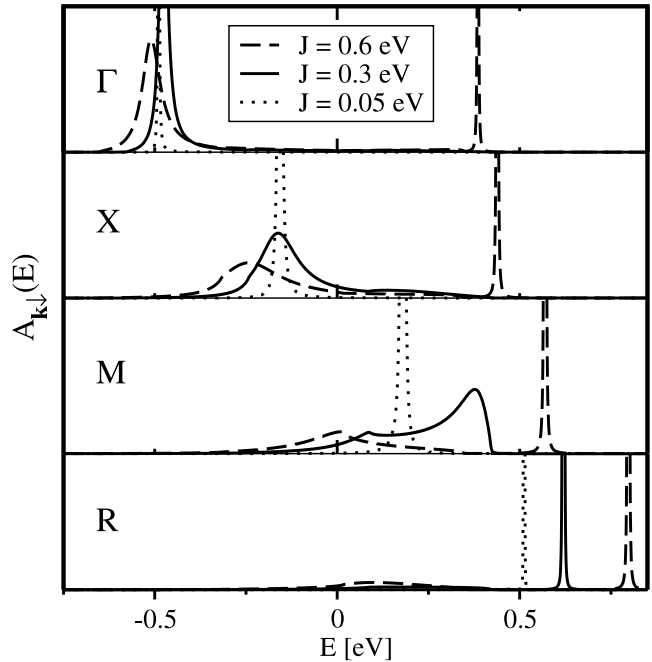}
    \caption{$\downarrow$-spectral density as function of energy for
      several symmetry points in the first Brillouin zone and for
      different exchange couplings $J$. Parameters:
      $S=\frac{1}{2},W=1eV$, sc lattice.}
    \label{fig:magnetic_polaron}
  \end{center}
\end{figure}
the parallel aligned spin system. Only the Ising-type interaction in
(\ref{eq:spinflip}) and (\ref{eq:flipflip}) takes care for a rigid shift of the total spectrum
by $-\frac{1}{2}JS$. The spectral density is a \mbox{$\delta$-function} at
$\epsilon(\mathbf{k})-\frac{1}{2}JS$ where $\epsilon$ is the free Bloch
energy, the Fourier transform of the hopping integral $T_{ij}$. Real
correlation effects appear, however, in the $\downarrow$
spectrum. Fig.~\ref{fig:magnetic_polaron}
shows 
the energy dependence of the $\downarrow$-spectral density for some
symmetry points. For weak couplings the spectral density consists of a single
pronounced peak. The finite width points to a finite quasiparticle lifetime due to
first spinflip processes, but the quasiparticle bandstructure,
being read off from the peak positions, is still very similar in shape
to the Bloch dispersion.

This changes drastically already for rather moderate effective exchange
couplings $JS/W$. One observes in certain parts of the Brillouin zone,
for a strongly coupled system even in the whole Brillouin zone, that the
excitation energy splits into two parts. The sharp high-energy peak
belongs to the magnetic polaron while the broader low-energy part
consists of scattering states due to magnon emission. As long as the
polaron peak is above the scattering spectrum the quasiparticle has even
an infinite lifetime. The scattering spectrum is in general rather broad
because the emitted magnon can carry away any wave-vector from the first
Brillouin zone. Because of the concomitant spinflip, magnon emission can
happen only if there are $\uparrow$ states within reach. Therefore,
the scattering part extends just over that energy region where
$\rho_{\uparrow}\neq 0$ ($\rho_{\sigma}$: quasiparticle density of
states (Q-DOS)).

For the general case (finite temperature, finite band occupation) the
many-body problem of the KLM cannot be solved exactly. Approximations
must be tolerated. The central quantity is the selfenergy, for which we
use the Green function procedure developed in \cite{NRMJ97}. This non-perturbational theory
can be considered a \textit{``moment conserving decoupling
  approximation''} (MCDA) which interpolates between exact limiting
cases as, e.~g., the above discussed example. In more detail, the method
decouples the hierarchy of equations of motion by expressing certain
``higher'' Green functions as linear combination of ``lower'' functions,
already involved in the procedure. The ansatz is motivated by exact
special cases as the atomic limit, ferromagnetic saturation, local spin
$S=1/2$, full or empty band, and so on. Free parameters in the ansatz
are eventually fitted to rigorous spectral moments. As other methods, too, e.~g.~the interpolating
selfenergy approaches in refs.\cite{nrrm01,nrrmk03}, it leads to the following structure of the selfenergy:
\begin{equation}
  \label{eq:selfenergy}
  \Sigma_{\mathbf{k}\sigma}(E)=-\frac{1}{2}Jz_{\sigma}\left<S^{z}\right>+J^{2}D_{\mathbf{k}\sigma}(E)
\end{equation}
Restriction to the first term, only, yields the mean-field approach to
the KLM, which is correct for sufficiently weak couplings $J$. It is
due to the Ising-part in Eq.~(\ref{eq:spinflip}) 
($z_{\sigma}=\delta_{\sigma\uparrow}-\delta_{\sigma\downarrow}$).
Without the second part of the selfenergy it would give rise to a
spin-polarized splitting of the conduction band. The term $D_{\mathbf{k}\sigma}(E)$ is more
complicated being predominantly determined by the spin exchange
processes. It is a complicated functional of the selfenergy itself,
i.~e.~(\ref{eq:selfenergy}) is an implicit equation for
$\Sigma_{\mathbf{k}\sigma}(E)$ and not at all an analytical
solution. $D_{\mathbf{k}\sigma}(E)$ depends, further on, on mixed spin
correlations such as
$\left<S_{i}^{z}n_{i\sigma}\right>$,$\left<S_{i}^{+}c_{i\downarrow}^{+}c_{i\uparrow}\right>$,
..., built up by combinations of localized-spin and itinerant-electron
operators. Fortunately, all these mixed correlations can rigorously be
expressed via the spectral theorem by any of the Green functions involved in
the hierarchy of the MCDA.

However, there are also pure local-moment correlations of the form
$\left<S_{i}^{z}\right>$, $\left<S_{i}^{\pm}S_{i}^{\mp}\right>$,
$\left<\left(S_{i}^{z}\right)^{3}\right>$, ..., which also have to be
expressed by the selfenergy (\ref{eq:selfenergy}). For this purpose we
use the \textit{modified} RKKY theory of~\cite{SaNo02} which exploits a
mapping of the interband exchange (\ref{eq:spinflip}) to an effective
Heisenberg model,
\begin{equation}
  \label{eq:Heisenberg}
  H_{f}=-\sum_{ij}\hat{J}_{ij}\mathbf{S}_{i}\cdot\mathbf{S}_{j}
\end{equation}
by averaging out the conduction electron degrees of freedom:
\begin{equation}
  \label{eq:mapping}
  -J\sum_{j}\mathbf{S}_{j}\cdot\mbox{\boldmath$\sigma$\unboldmath}_{j}
  \longrightarrow-J\sum_{j}\mathbf{S}_{j}
  \cdot\left<\mbox{\boldmath$\sigma$\unboldmath}_{j}\right>^{(c)}\longrightarrow H_{f}
\end{equation}
In the last analysis this means to determine the expectation value
$\left<c_{\mathbf{k}+\mathbf{q}\sigma}^{+}c_{\mathbf{k}\sigma^{'}}\right>^{(c)}$.
We use again a Green-function procedure \cite{SaNo02} which eventually leads to
effective exchange integrals in (\ref{eq:Heisenberg}):
\begin{eqnarray}
  \label{eq:exchangeintegral}
  \hat{J}_{ij}=\frac{J^{2}}{4\pi N^{2}}\sum_{\mathbf{kq}
    \sigma}e^{i\mathbf{q}\cdot(\mathbf{R}_{i}-\mathbf{R}_{j})}
  &&\int_{-\infty}^{+\infty}dE\:f_{-}(E)\textrm{Im}\left[\left(E-\epsilon(\mathbf{k})+i0^{+}\right)
  \right.\nonumber\\
  &&\times\left.\left(E-\epsilon(\mathbf{k+q})-\Sigma_{\mathbf{k}+\mathbf{q}\sigma}(E)\right)\right]^{-1}
\end{eqnarray}
$f_{-}(E)$ denotes the Fermi function. The effective exchange integrals
are decisively influenced by the conduction electron selfenergy
$\Sigma_{\sigma}$ which brings a distinct band occupation and temperature
dependence to the $\hat{J}_{ij}$. Neglecting $\Sigma_{\sigma}$ leads to
the ``conventional'' RKKY formula with $J_{ij}\propto J^{2}$ as a result
of second order perturbation theory. Via $\Sigma_{\sigma}$ higher order
terms of the electron spin polarization enter the
\textit{modified} RKKY being therefore not restricted to weak
couplings, only. Note that at the right-hand side of
Eq.~(\ref{eq:exchangeintegral}) the product of two propagators
appear. From the equation of motion method in \cite{SaNo02} it follows
directly that one of them is dressed by a selfenergy, the other not.

To get from the effective operator (\ref{eq:Heisenberg}) the magnetic
properties of the KLM we apply the standard Tyablikov-approximation
which is known to yield convincing results in the low as well as in the
high temperature region \cite{SaNo02}. All the above-mentioned
local-moment correlations are then expressed by the electronic
selfenergy. We therefore end up with a closed system of equations that can
be solved self-consistently for all entities of interest. 

\begin{figure}[htbp]
  \begin{center}
    \includegraphics[width=0.8\linewidth]{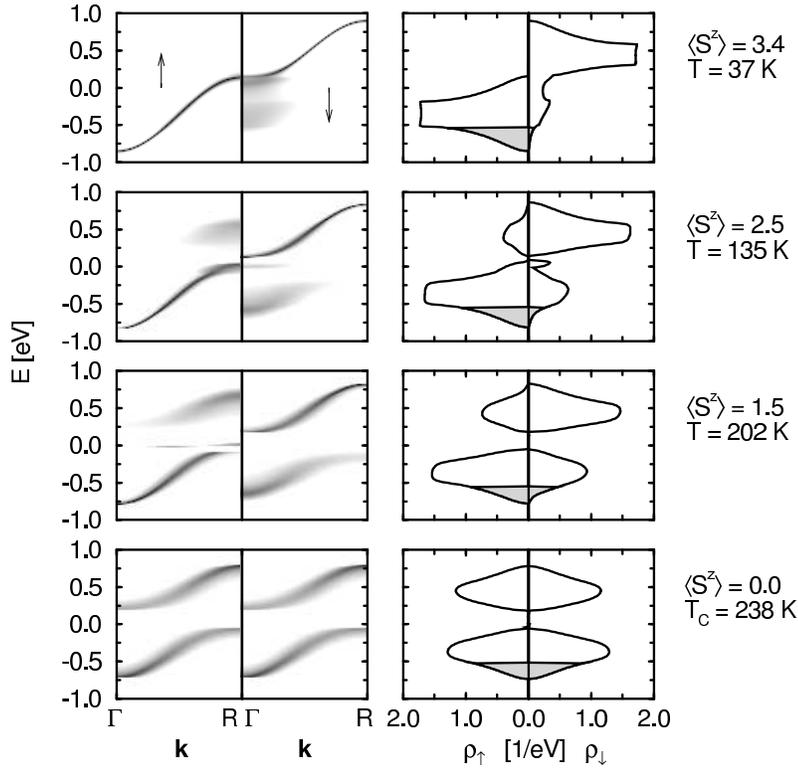}
    \caption{Quasiparticle bandstructure as a function of wave-vector
      (left column) and quasiparticle density of states as a function of
      energy (right column) for four different temperatures calculated
      within the MCDA\cite{NRMJ97}. Parameters: $J=0.2$~eV, $W=1$~eV, $n=0.2$, $S=\frac{7}{2}$,
      sc lattice.}
    \label{fig:QDOS-QBS}
  \end{center}
\end{figure}

Fig.~\ref{fig:QDOS-QBS} shows the temperature dependence of the
quasiparticle bandstructure (Q-BS), derived as density plot from the
spectral density, and the quasiparticle density of states
(Q-DOS) $\rho_{\sigma}(E)$ for a typical parameter set from the moderate
coupling regime ($J=0.2$~eV, $W=1$~eV, $n=0.2$, $S=\frac{7}{2}$, sc lattice). In
this case the self-consistently calculated Curie temperature $T_{\rm{C}}$
amounts to $238$~K. At $T=37$~K 
the local-moment magnetization is $3.4$
and therewith very close to saturation. The theory reproduces, in spite
of the finite carrier concentration, the same features as for the exact
$(n=0, T=0)$-case exhibited in Fig.~\ref{fig:magnetic_polaron}. The
$\uparrow$ dispersion as well as the QDOS
$\rho_{\uparrow}$ are practically identical to the respective ``free''
Bloch functions. The
$\downarrow$ spectrum, however, is more complicated. The scattering states take away a substantial part
of the spectral weight near the $\Gamma$ point while near the $R$ point 
the polaron state clearly dominates. With increasing temperature (decreasing $\left<S^z\right>$) a finite
magnon density appears allowing for scattering states in the $\uparrow$
spectrum, too, because of magnon absorption and simultaneous spinflip by
the $\uparrow$ electron. Furthermore, polaron
and scattering states separate, where, surprisingly, the rather broad
scattering spectrum is in wide wave-vector regions bunched to a
prominent peak. At $T=T_{\rm{C}}$ the spin asymmetry is removed but there
remains a correlation-caused splitting of the spectrum into two branches
and two subbands, respectively, due to the interband exchange
coupling $J$. Because of the possibility of mutual spin exchange
$\rho_{\uparrow}$ and $\rho_{\downarrow}$ are occupying for finite
temperatures always the same energy regions.

The key-quantity of ferromagnetism is the Curie temperature $T_{\rm{C}}$. A
finite $T_{\rm{C}}$ comes out as a consequence of an indirect coupling between
the local moments, mediated by a polarization of the conduction electron
\begin{figure}[htbp]
  \begin{center}
    \includegraphics[width=0.5\linewidth]{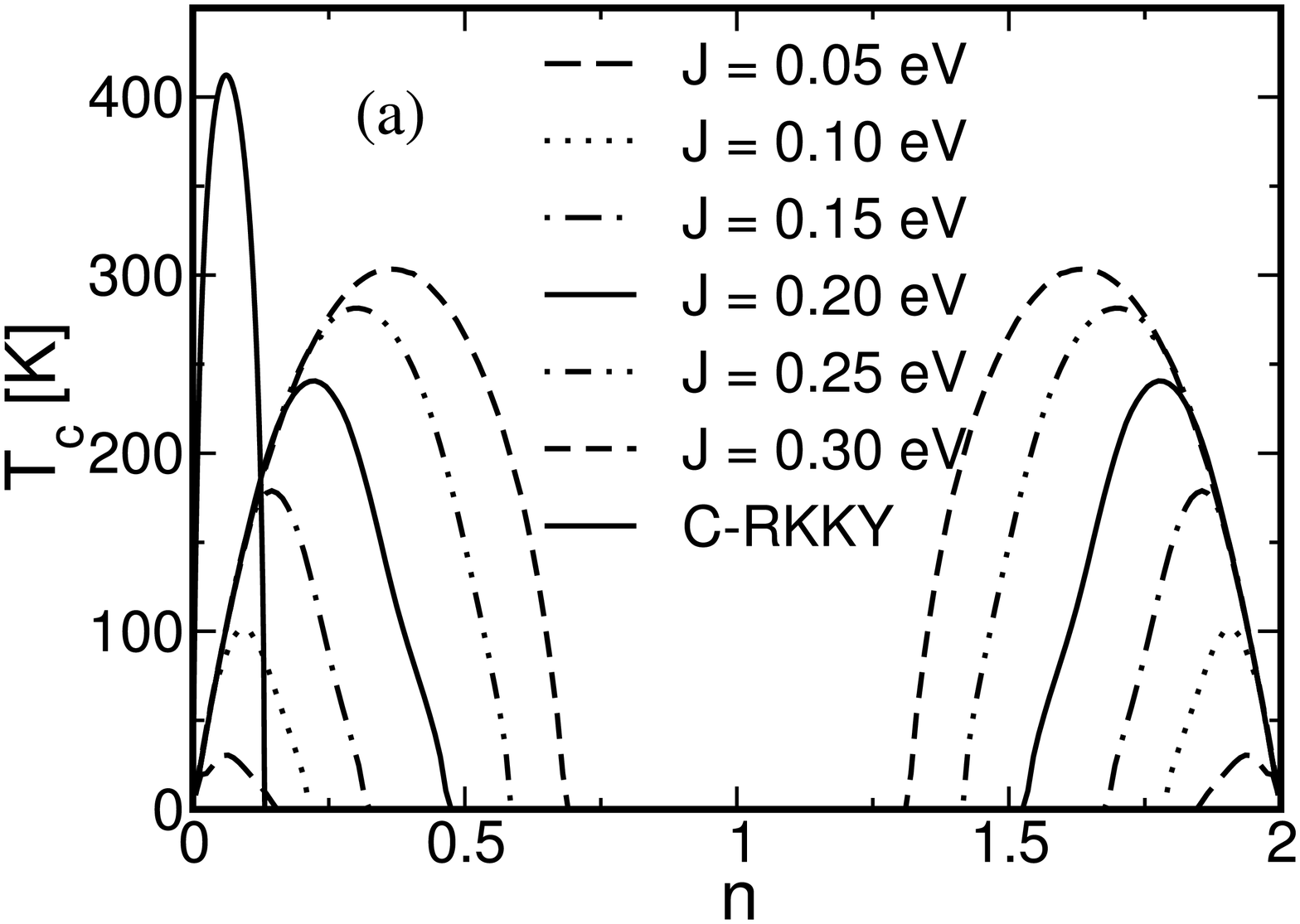}\hfill
    \includegraphics[width=0.5\linewidth]{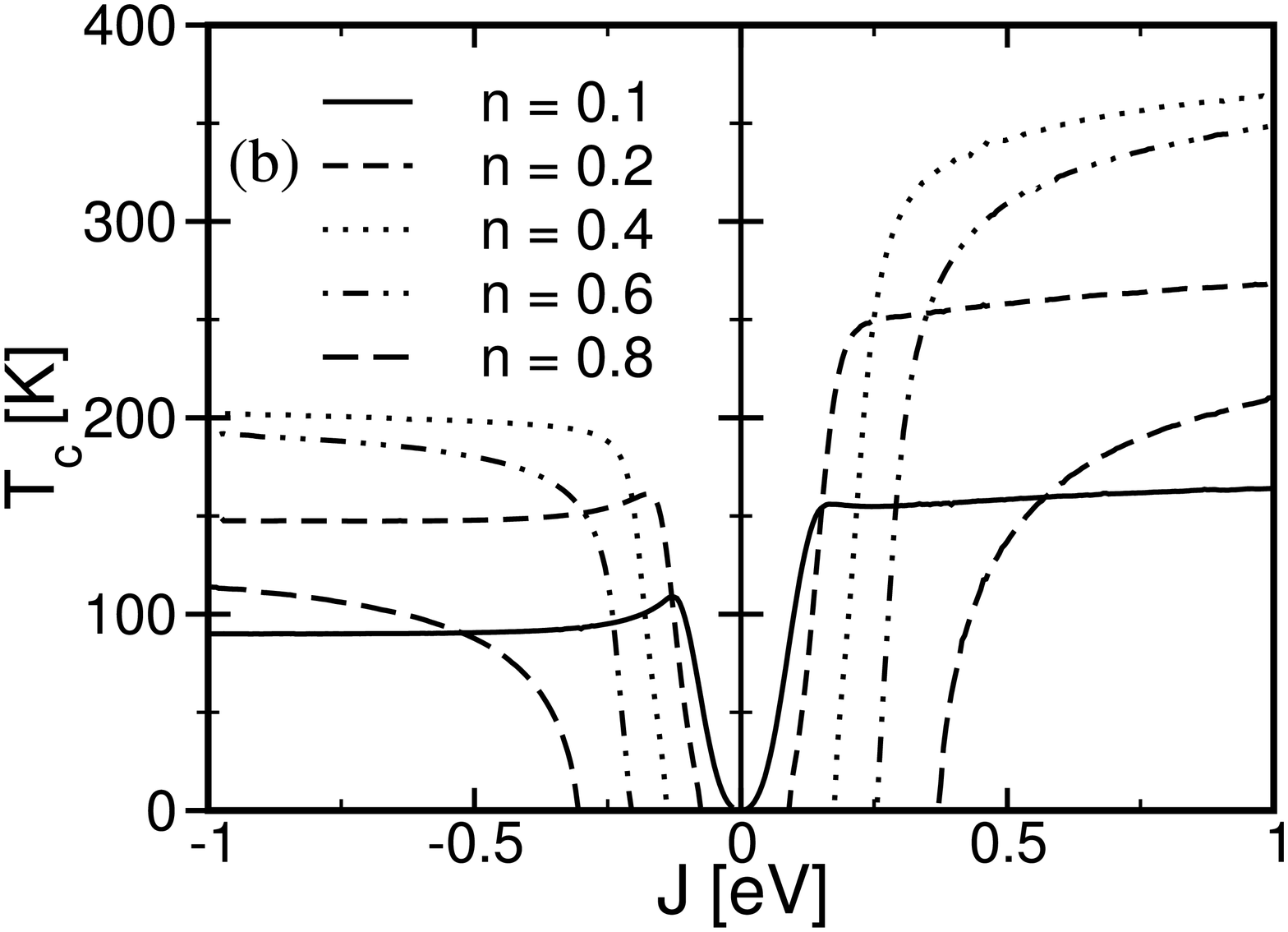}
    \caption{Curie temperature as a function of (a) the band 
      occupation $n$ for various $J$, (b) the interband exchange 
      coupling $J$ for various $n$, calculated by use of the \textit{modified} RKKY.}
    \label{fig:Curietemperature}
  \end{center}
\end{figure}
spins. Therefore, $T_{\rm{C}}$ exhibits a strong carrier-concentration
dependence which is plotted in
Fig.~\ref{fig:Curietemperature}. The effective exchange integrals have
been taken into account up to 25th neighbours. Ferromagnetism appears for low electron
(hole) concentrations, while being excluded in the region around
half-filling ($n=1$). Around $n=1$ we expect an antiferromagnetic phase,
however, not explicitly calculated here. A similar $n$-dependence of $T_{\rm{C}}$ has been found in
Ref.~\cite{ChaMil01}. It is interesting to compare the results with
those of the \textit{conventional} RKKY, given by $\Sigma_{\sigma}\equiv
0$ in Eq.~(\ref{eq:exchangeintegral}). A corresponding curve is inserted in
Fig.~\ref{fig:Curietemperature}. The maximal $T_{\rm{C}}$ values are higher, but
ferromagnetism exists only in a very narrow region of low electron
(hole) concentrations where this region turns out to be independent of $J$.

Two features dominate the $J$-dependence of the Curie temperature plotted
in part (b) of Fig.~\ref{fig:Curietemperature}. The first is the
appearance of a critical $J$ for band occupations for which the
\textit{conventional} RKKY does not allow ferromagnetism ($n\ge
0.13$). This is in
accordance with the fact that for $J\rightarrow 0$ the \textit{modified} RKKY
reproduces the \textit{conventional} RKKY. The second fact is that $T_{\rm{C}}$
runs into a saturation for strong couplings $J$, where the saturation
value depends on the band occupation $n$. These results are far beyond
\textit{conventional} RKKY which can work of course only in the weak
coupling regime.

\begin{figure}[htbp]
  \begin{center}
    \includegraphics[width=0.485\linewidth]{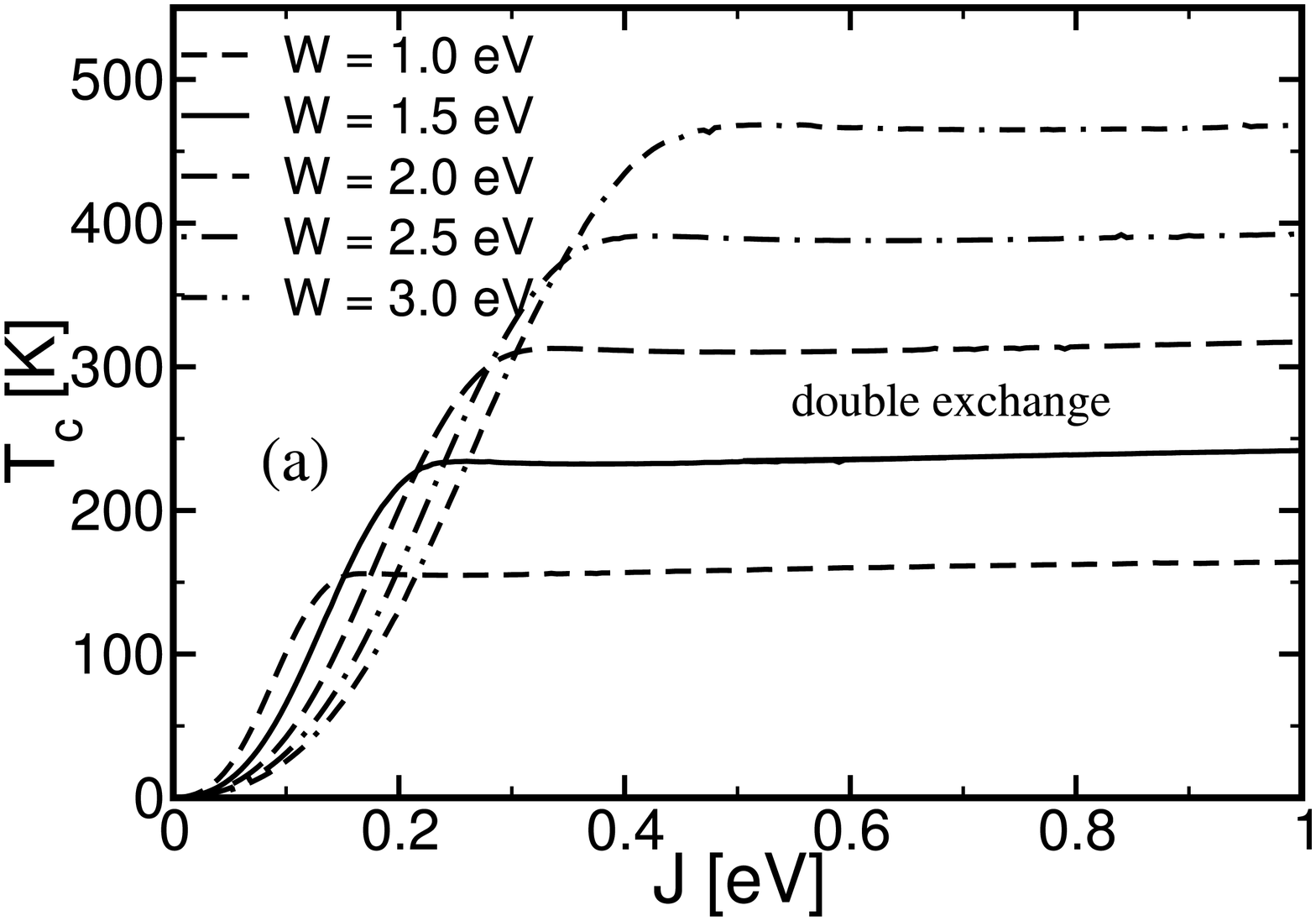}\hfill
    \includegraphics[width=0.485\linewidth]{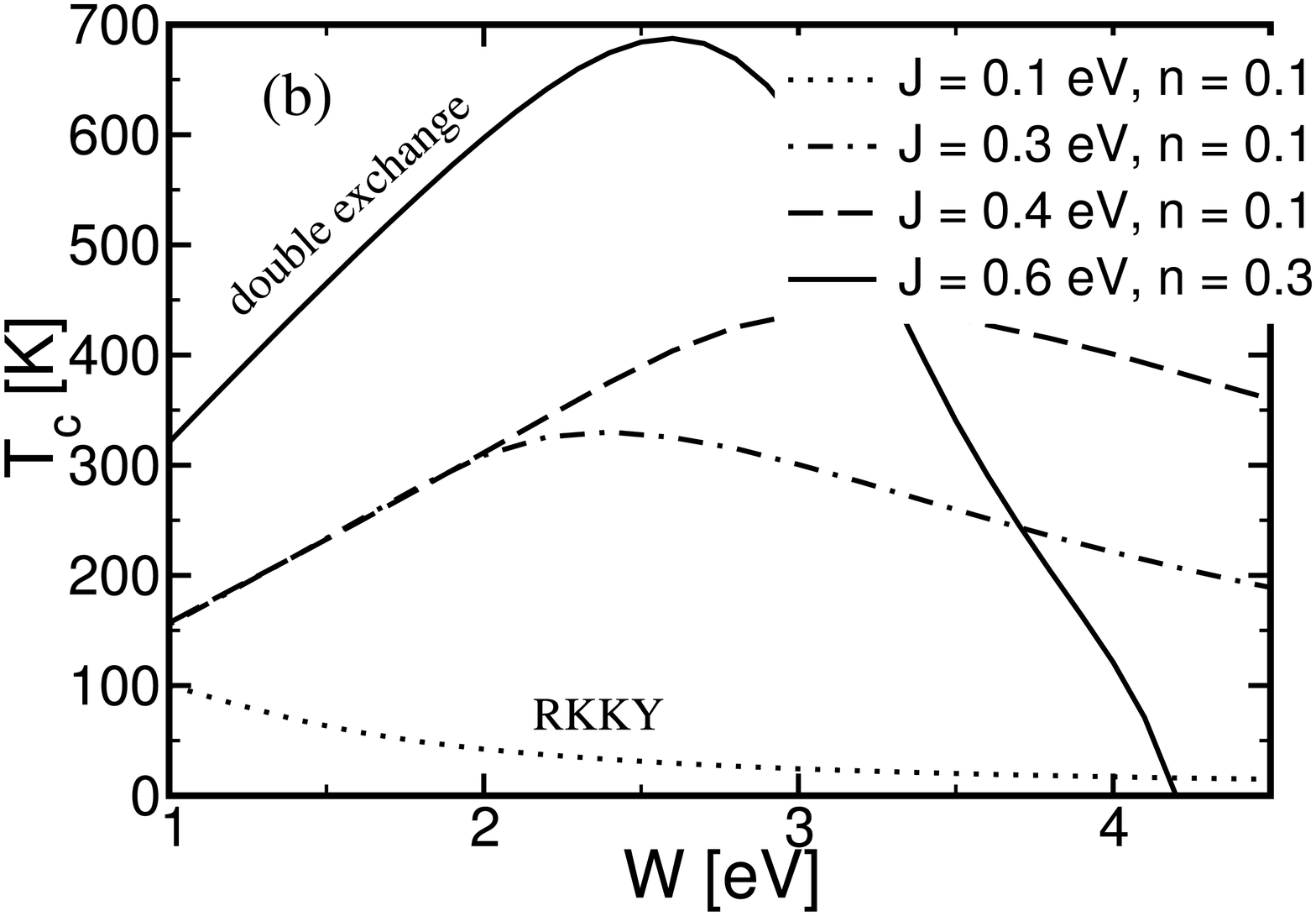}
    \caption{Curie temperature as a function of (a) the interband
      exchange $J$ for various $W$ and $n=0.1$, (b) the Bloch bandwidth
      $W$ for various 
      $J$ and $n$, calculated by use of the \textit{modified} RKKY.}
    \label{fig:doubleexchange}
  \end{center}
\end{figure}

It is interesting to observe (see Fig.~\ref{fig:doubleexchange}) that in
the weak coupling RKKY \textit{region} $T_{\rm{C}}$ scales with the effective
coupling constant $\frac{J}{W}$ and in the strong coupling
\textit{double exchange region} with 
the kinetic energy, i.~e.~$\propto
W$. In part (b) of Fig.~\ref{fig:doubleexchange} the change from RKKY
behaviour (small $\frac{J}{W}$) to double exchange behaviour (large
$\frac{J}{W}$) is clearly to be seen.

\begin{figure}[htbp]
  \begin{center}
    \includegraphics[width=0.6\linewidth]{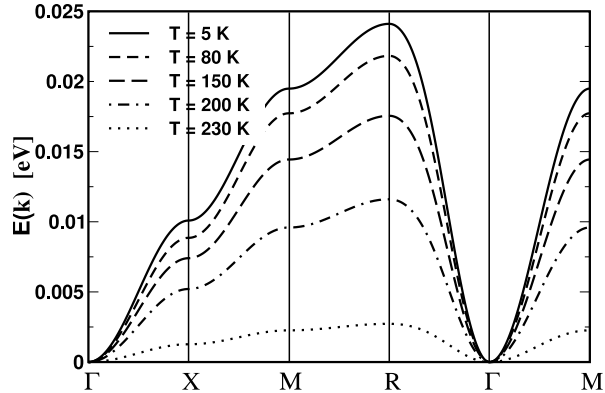}
    \caption{Spin-wave dispersion of the ferromagnetic Kondo-lattice model as a
      function of the wave-vector for different temperatures $T$. Parameters: $J=0.2$~eV,
      $n=0.2$, $S=7/2$, sc lattice, and $W=1$~eV. The self-consistently calculated
      Curie temperature is $T_{\rm{C}}=232$~K.}
    \label{fig:spin_waves}
  \end{center}
\end{figure}

Fig.~\ref{fig:spin_waves} shows a typical example of a spin-wave dispersion, which obtains
via the effective exchange integrals (\ref{eq:exchangeintegral}) a
distinct temperature dependence. Upon heating the dipersion relation
uniformly softens, disappearing above $T_{\rm{C}}$. This agrees qualitatively
with neutron scattering data on manganites \cite{hwang98}.

\section{Application to ferromagnetic EuO and EuS}
In the last step we want to demonstrate how the model analysis can be used to
describe real magnetic materials. We exemplify the method with respect to the classical
local-moment ferromagnets EuO and EuS. These are ferromagnetic semiconductors
which crystallize in the fcc-rocksalt structure with lattice constants
$a=5.142$~\AA (EuO) and $a=5.95$~\AA (EuS), respectively. The physics of these
materials is mainly determined by localized and half-filled 4f levels, on the
one hand, and, on the other hand, by extended, at low temperature empty, 5d band states. In
accordance to Hund's rule, the Eu$^{2+}$-4f shell produces a maximum spin moment
$S=\frac{7}{2}$.

To investigate real materials such as EuO and EuS one has to exploit the
multiband version of the KLM developed in Section 1. As already explained, the
hopping integrals $T_{ij}^{m\bar{m}}$ in (\ref{eq:kinEn}) have not only to
incorporate the kinetic energy of the band electrons plus the influence of the
\begin{figure}[htbp]
  \begin{center}
    \includegraphics[width=0.7\linewidth]{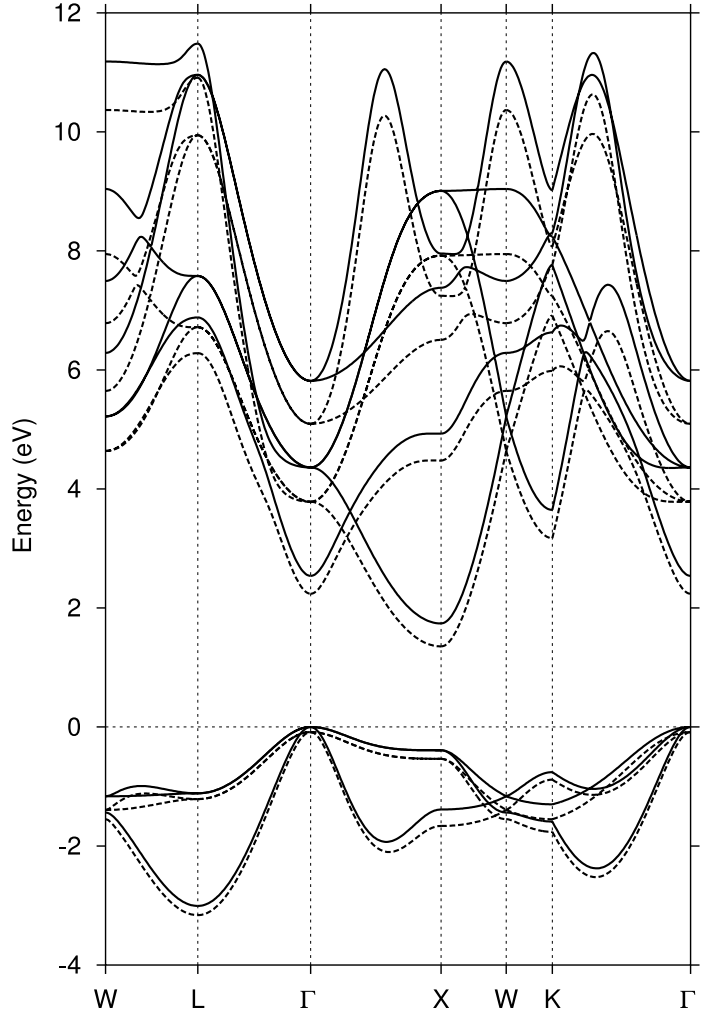}
    \caption{Spin-dependent (solid lines down-spin, broken lines up-spin)
      band structure of bulk EuS calculated within a TB-LMTO scheme with the 4f levels
      treated as core states. The energy zero coincides with the Fermi energy.}
    \label{fig:blkbsEuS}
  \end{center}
\end{figure}
periodic lattice potential but also as realistically as possible all those
interactions which are not directly covered by the model Hamiltonian. For this
purpose we performed a bandstructure calculation based on density functional
theory, using the Andersen scheme \cite{andersen75,andersen84} of a \textit{tight-binding linear
muffin-tin orbital} (TB-LMTO) ansatz. LDA-typical difficulties arise with the
strongly localized character of the 4f levels. A \textit{normal} LDA calculation
for EuO produces a metal with the 4f levels lying well within the conduction
band. To circumvent the problem we considered the 4f electrons as core electrons, since
our main interest is focused on the temperature-reaction of the (empty) conduction
bands. In our study the 4f levels appear only as localized spins (moments) in the sense
of $H_{ff}$ in Eq.~(\ref{eq:filmheisenberg}). Fig.~\ref{fig:blkbsEuS}
shows the 
so-derived spin-dependent bandstructure of EuS
without the 4f levels. The low-energy part belongs to S-3p states, we are not
interested in here, because only $fd$ excitations shall be in the
focus. The conduction band region is dominated by Eu-5d states, to which
we have restricted the investigation by our combined many-particle, \textit{first
principles} procedure. For comparison we have also performed an LDA+U
calculation which indeed 
is able to put the respective bands into better positions. On the other hand, the
results, which we are interested in, do not differ too strongly from
those of LDA 
with the 4f electrons in the core.
So we have chosen the much simpler LDA calculation, since our study mainly aims at
overall correlation and temperature effects. The extreme details of the band structure
are not so important.

The many-body treatment of the multiband KLM follows exactly the same line as for the
single-band case, no need to create new approaches. However, we have to fix the only
parameter of our procedure, the interband exchange $J$. It is generally assumed that an
LDA treatment of ferromagnetism is quite compatible with a mean-field picture, so that
the $T=0$-shift of the $\uparrow$ and $\downarrow$ dispersions in Fig.~\ref{fig:blkbsEuS} should
amount to $\Delta=JS$. However, it can be seen
that the assumption of a rigid shift is too simple. A certain energy dependence of the
exchange splitting is found by LDA, too. We decided to use the splitting of the 5d-band
centre of gravity to fix the $J$ for EuS at $0.23$~eV \cite{MNO02}.
Averaging over prominent features in the Q-DOS of EuO yields a similar
value of $J=0.25$~eV \cite{SNO01}.

Another important issue concerns the renormalized single-particle input, because a
double counting of the decisive interband exchange, once by the LDA input and once more
explicitly in the model-Hamiltonian (\ref{eq:spinflip}), might produce rather misleading results. The
most direct solution of this problem would be to switch off the interband exchange
$H_{df}$ in the LDA code, which turns out to be impossible. However we can elegantly
exploit the exact limiting case, discussed at the beginning of Section 3. For zero
temperature and zero band occupation the $\uparrow$ spectrum of the single- as well as
multiband-KLM is only rigidly shifted towards lower energies by the amount of
$\frac{1}{2}JS$. Thus we can use without any manipulation the $\uparrow$ dispersions of the LDA calculation
($T=0 !$) for the empty 5d bands as input for the single-electron Hamiltonian
(\ref{eq:kinEn}).
There is no need to switch off $H_{df}$ because in this special case it leads only to
an unimportant rigid shift. Note that this holds only for the $\uparrow$ part, the
$\downarrow$ spectrum is even at $T=0$ strongly influenced by correlation effects (see
Fig.~\ref{fig:magnetic_polaron}). Therewith it is guaranteed, on the one side, that all those interactions, which
are not directly covered by the KLM, are implicitly taken into account by the
LDA-renormalized single-particle energies. On the other side, a double counting of any
decisive interaction, which usually occurs when combining first-principles and
many-body model calculations, is definitely avoided. 

Because of the empty conduction bands a self-consistent justification of the EuO (EuS)
ferromagnetism via the modified RKKY-treatment of Section 3 is not possible. The
magnetic properties of the insulators EuO and EuS are exclusively due to the partial
operator $H_{ff}$ (\ref{eq:filmheisenberg}). Its exchange integrals between nearest ($J_{1}$) and
next-nearest ($J_{2}$) neighbours have been derived by a neutron scattering spin wave
analysis \cite{BDK80}: $J_{1}/k_{\rm{B}}=0.625 (0.221)$~K,
$J_{2}=0.125 (-0.100)$~K for EuO (EuS). The single-ion anisotropy constant $D$ does not
play a decisive role as long as the bulk bandstructure and its temperature dependence
is aimed at. However, when treating systems of lower dimensionality (films, surfaces)
as done for EuO in Refs.~\cite{schiller01:_predic_euo,schiller01:_kondo}, then a
finite $D$ is necessary to overcome the Mermin-Wagner theorem
\cite{gelfert01} for getting a collective magnetic order of the moment
system. $D$ has been
chosen in such a way that the experimental $T_{\rm{C}}$ value is correctly reproduced (see
Fig.~\ref{fig:Curietemperature} in Ref.~\cite{MNO02}. The temperature-dependence
of the 4f moment
system leads to a remarkable induced temperature behaviour of the (unoccupied!) 5d
spectrum. The evaluation  of the effective Heisenberg model
(\ref{eq:filmheisenberg}) is done by a Tyablikov
decoupling of the equation of motion of a poperly chosen spin Green
function. Details are given in \cite{SaNo02,NRMJ97}.

Let us now select some typical results found for the archetypal ferromagnetic
semiconductors EuO and EuS. We have no intention to be complete in this overview,
rather to demonstrate the possibilities of our combined many-body/\textit{ab initio}
\begin{figure}[htbp]
  \begin{center}
    \includegraphics[width=0.6\linewidth]{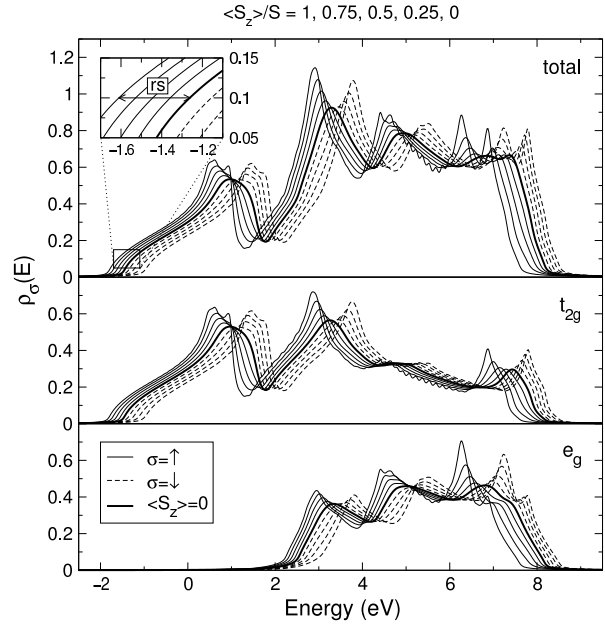}
    \caption{Temperature-dependent densities of states of the Eu-5d bands of
      bulk EuO ($J=0.25$~eV). Solid lines: $\uparrow$-spectrum; broken lines:
      $\downarrow$-spectrum. Thick line: spectra for $T=T_{C}$ ($\langle
      S^{z}\rangle=0$). The outermost curves belong to $T=0$ ($\langle S^{z}\rangle=S$).
      With increasing temperature the $\uparrow$ and $\downarrow$ curves approach each
      other.}
    \label{fig:qdosEuO}
  \end{center}
\end{figure}
%%%%%%%%%%%%%%%%%%%%%%%%%%%%%%%%%%%%%%%%%%%%%%%%%%%%%%%%%%%%%%%
\begin{figure}[htbp]
  \begin{center}
    \includegraphics[width=0.6\linewidth]{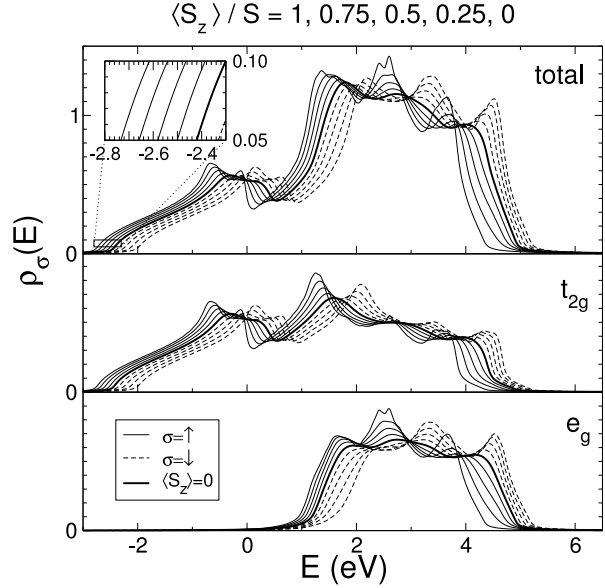}
    \caption{The same as in Fig.~\ref{fig:qdosEuO} but for bulk EuS ($J=0.23$~eV).}
    \label{fig:qdosEuS}
  \end{center}
\end{figure}
method. Fig.~\ref{fig:qdosEuO} and \ref{fig:qdosEuS} show the
quasiparticle density 
of states (Q-DOS) for EuO and EuS,
respectively, for five different 4f magnetizations, i.~e.~five different temperatures.
Not surprisingly, the Q-DOS of EuO (Fig.~\ref{fig:qdosEuO}) and EuS 
(Fig.~\ref{fig:qdosEuS}) are qualitatively rather
similar. Since we have taken into account the full bandstructure of the Eu-5d
conduction bands, the symmetry of the different 5d orbitals is preserved. The 5d bands
can therefore be decomposed into $t_{2g}$ and $e_{g}$ subbands, where the
$t_{2g}$ bands are
substantially broader ($\sim 7$~eV (EuS), $\sim 9$~eV (EuO)) than the $e_{g}$
bands ($\sim$ 4~eV
(EuS), $\sim$ 6~eV (EuO)). A remarkable temperature-dependence shows up for both
ferromagnets, manifesting itself, e.~g., in a shift of the lower edge of the $\uparrow$
Q-DOS to lower energies upon cooling from $T=T_{\rm{C}}$ to $T=0$. This explains the famous
red shift of the optical absorption edge for the electronic $4f-5d_{t_{2g}}$
transition, in the meantime observed for all ferromagnetic local-moment systems
\cite{batlogg75}. The temperature behaviour do not comply with the
simple Stoner picture of an energy-independent induced exchange splitting of the
spectra. The reason for the observed more complicated behaviour as a function of
temperature is that in our theory correlation is treated in a way distinctly beyond
mean-field.

\begin{figure}[htbp]
  \begin{center}
    \includegraphics[width=0.6\linewidth]{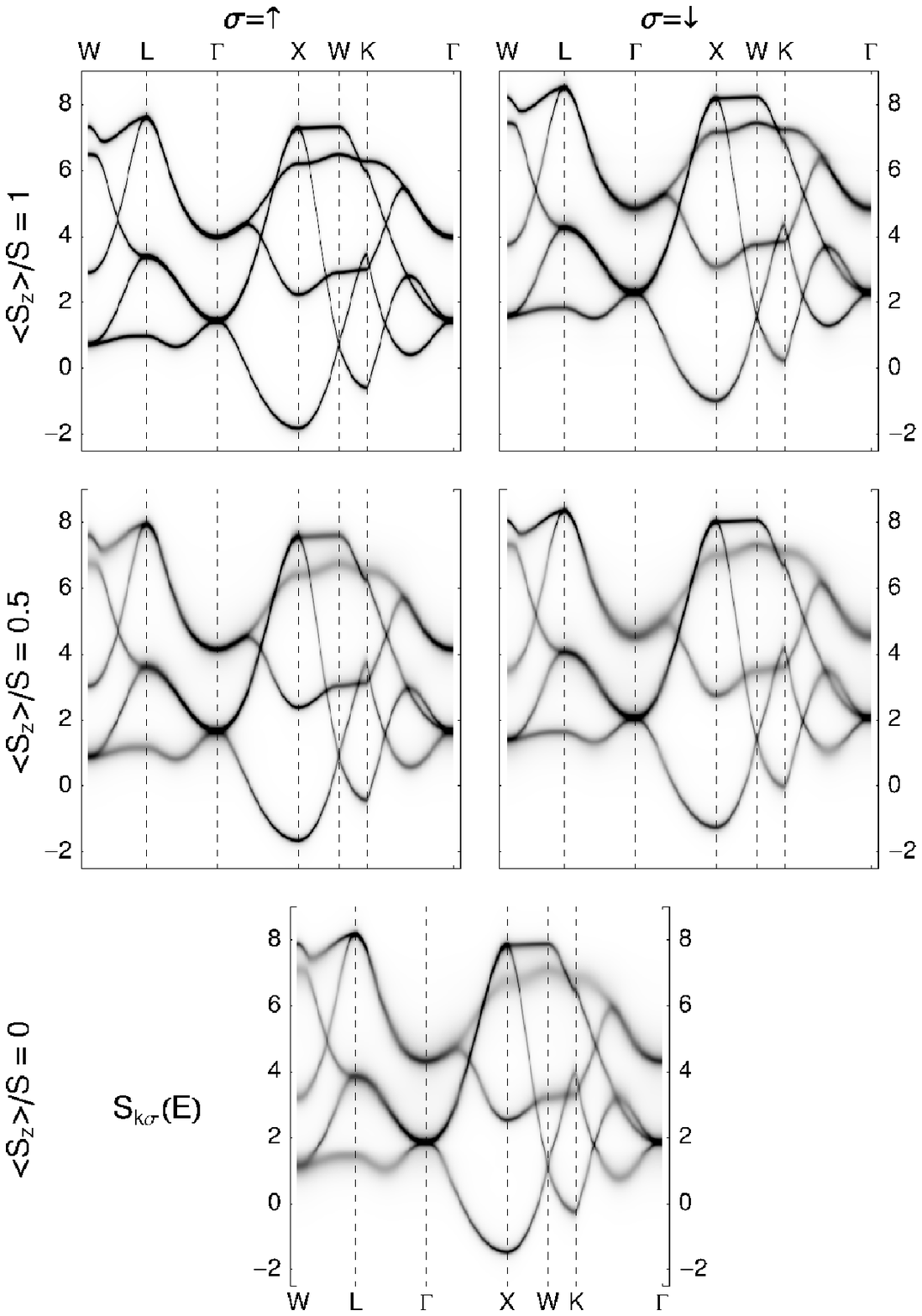}
    \caption{Spin-dependent quasiparticle band structure of the 5d-bands of
      bulk EuO ($J=0.25$~eV) for three different 4f magnetizations $\langle
      S^{z}\rangle$.}
    \label{fig:spin_qdos_EuO}
  \end{center}
\end{figure}

While the Q-DOS refers to the spin resolved but angle averaged, direct or inverse
photoemission experiment, the $\mathbf{k}$ dependent spectral density is the angle
resolved counterpart. Prominent peaks of the spectral density as a function of energy
determine the $E(\mathbf{k})$ quasiparticle bandstructure (Q-BS). As an example,
Fig.~\ref{fig:spin_qdos_EuO} represents the EuO-spectral density as a
density plot for some symmetry directions and
%%%%%%%%%%%%%%%%%%%%%%%%%%%%%%%%%%%%%%%%%%%%%%%%%%%
three different temperatures. The degree of blackening is a measure of the magnitude of
the spectral density. For $T=0$ ($\left<S_{z}\right>=S$) the $\uparrow$ spectral density
agrees, except for a constant shift, with the LDA result, being therefore a
$\delta$-function at the excitation energy, corresponding to a quasiparticle with
infinite lifetime. The $\downarrow$ spectral density, however, exhibits already  at $T=0$
a broadening of the dispersion curves, most notably around the $\Gamma$-point, which
indicates finite quasiparticle lifetimes due to correlation effects. For intermediate
temperatures ($\left<S_{z}\right>=0.5\;S$) a broadening sets in also  for the $\uparrow$
dispersions because the $\uparrow$ electron, too, can now exchange its spin with
the no
longer saturated local moment system. In the $\downarrow$ spectrum correlation effects
increase leading to further quasiparticle damping. Simultaneously, both spin spectra
are shifted towards each other reducing therewith the effective exchange splitting. At
$T=T_{\rm{C}}$ ($\left<S_{z}\right>=0$) $\uparrow$ and $\downarrow$ dispersions conincide.
Very similar behaviour is found for EuS \cite{MNO02}.

\begin{figure}[htbp]
  \begin{center}
    \includegraphics[width=0.5\linewidth]{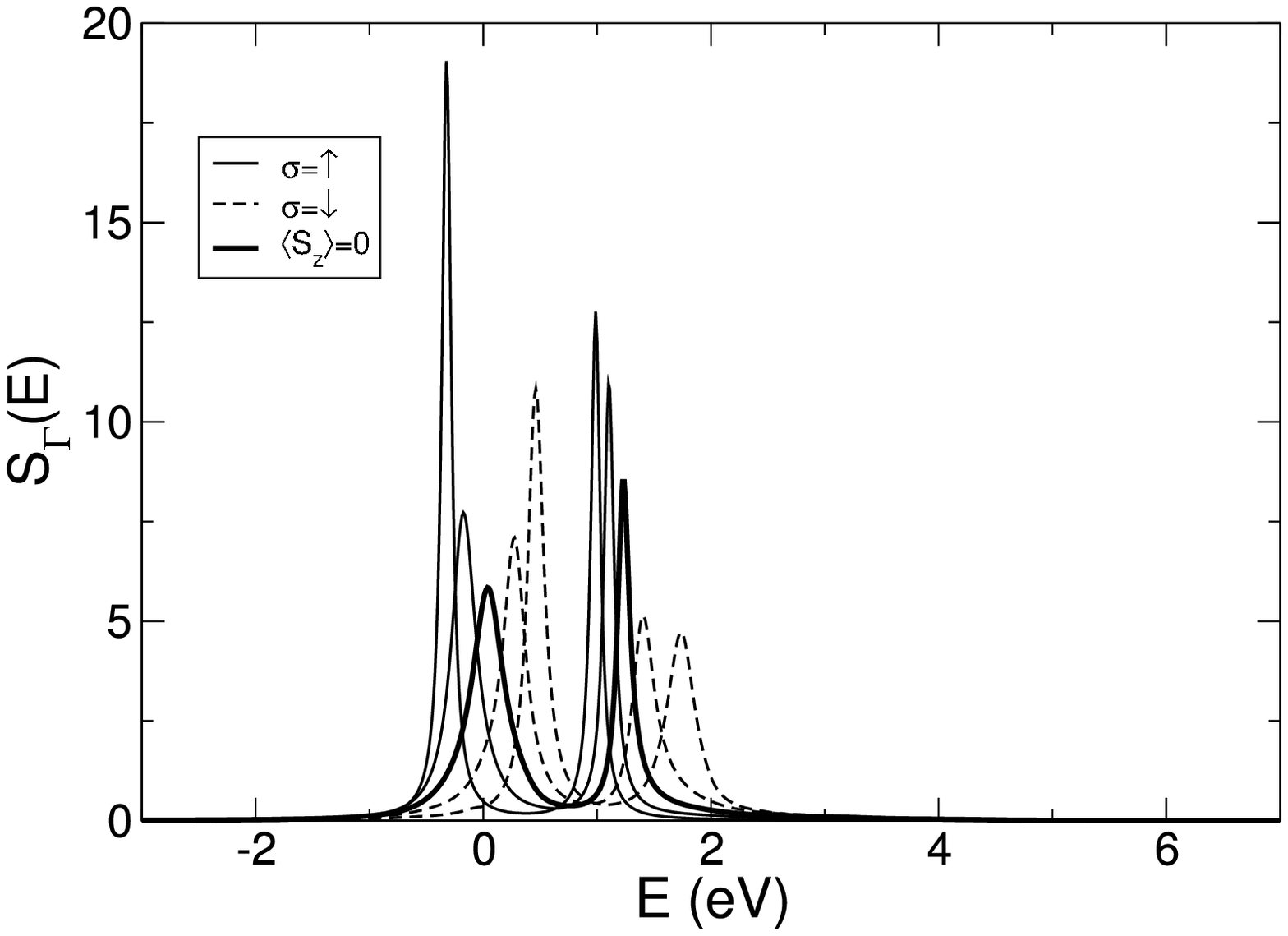}\hfill
    \includegraphics[width=0.5\linewidth]{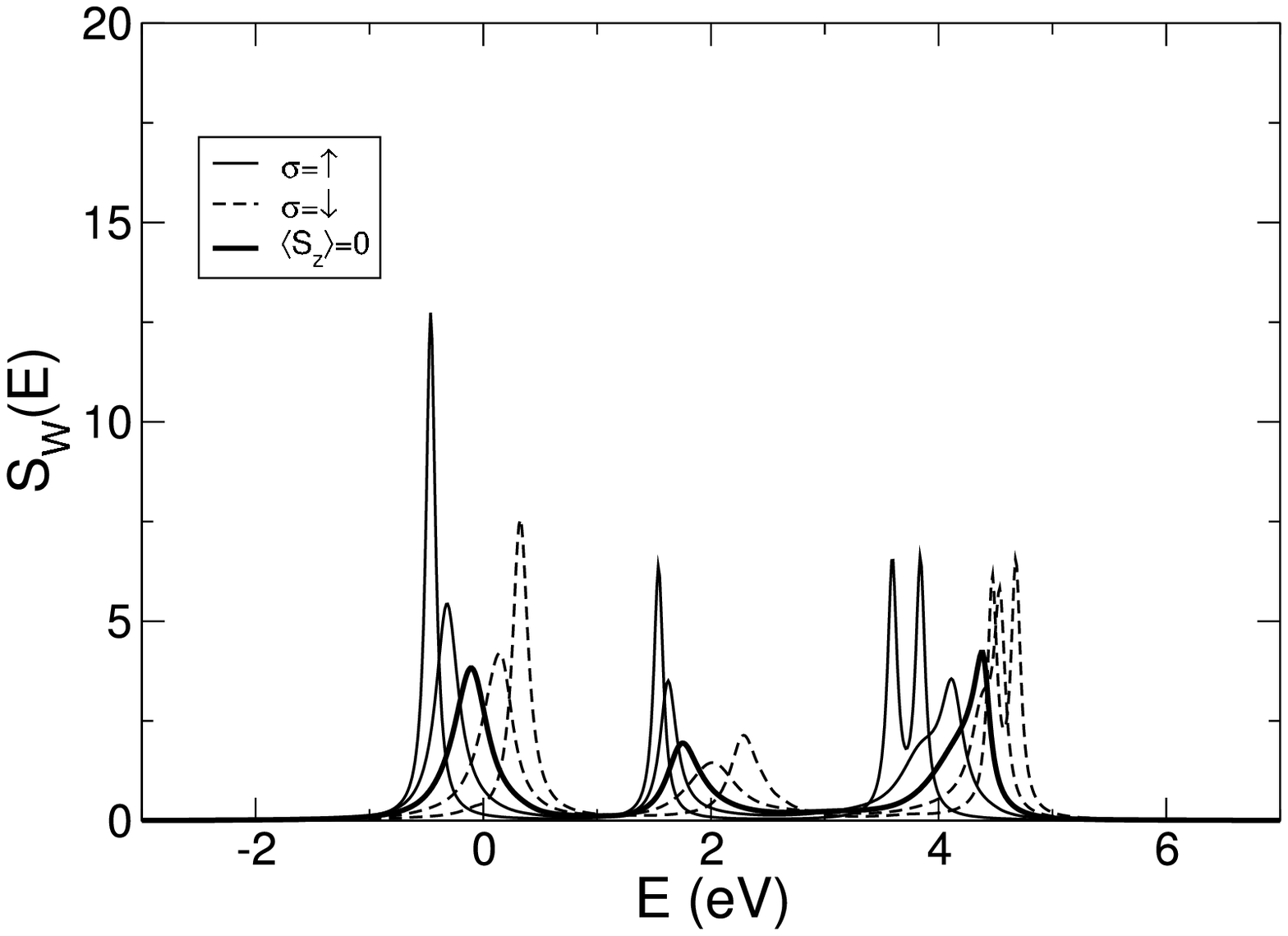}
      \caption{(a) Spin-dependent spectral density $S_{\mathbf{k}\sigma}$ for
      $\mathbf{k}$ from the $\Gamma$-point (5d states) of bulk EuS ($J=0.23$~eV) as a
      function of the energy for three different 4f magnetizations: $\langle
      S_{z}\rangle/S=1.0, 0.5, 0$. Thick line for $T=T_{\rm{C}}$. Solid lines: up-spin:
      broken lines: down-spin. Curves with maximum spin splitting of the peaks belong
      to $T=0$. (b) The same as in (a) but for the W-point.}
    \label{fig:SkGWEuS}
  \end{center}
\end{figure}

As mentioned, the Q-BS has been derived from the spectral density. To give an insight
into the temperature behaviour of the spectral density we have plotted in Fig.~\ref{fig:SkGWEuS} for two
$\mathbf{k}$-points from the first Brillouin zone ($\Gamma$, W) the energy dependence of
the spectral density for the same three 4f magnetizations as in Fig.~\ref{fig:spin_qdos_EuO}, but now for
EuS. For the $\Gamma$-point (Fig.~\ref{fig:SkGWEuS}~(a)) we have to expect two structures
according to the twofold degenerate ($e_{g}$) and threefold degenerate
($t_{2g}$) 
dispersions. Well defined quasiparticle
peaks appear, spin split below $T_{\rm{C}}$. The spin splitting, induced via the interband
coupling to the magnetically active 4f system, collapses for $T\rightarrow T_{\rm{C}}$. One
speaks of a \textit{Stoner-type behaviour}. Quasiparticle damping, that scales with the
peak width, obviously increases with temperature. At the W-point (Fig.~\ref{fig:SkGWEuS}~(b)) four
sharp peaks show up in the $\uparrow$ spectrum, and, though already strongly damped,
the same peak sequence appears in the $\downarrow$ part, too (compare with the Q-BS in
Fig.~\ref{fig:spin_qdos_EuO}). The reduction of the spin splitting with increasing temperature leads to a
strong overlap of the two upper peaks, which are then no longer distinguishable.
%%%%%%%%%%%%%%%%%%%%%%%%%%%%%%%%%%%%%%%%%%%%%%%%%%%%%%%%%%%%%%%%
Altogether, the 5d spectral densities of the ferromagnetic semiconductors EuS and EuO
(not shown here) exhibit drastic temperature-dependencies, what concerns the positions
and the widths of the quasiparticle peaks. The driving force is the interband (4f-5d)
exchange interaction, the investigation of which was the main aim of the presented study.

\section{Summary}

We have demonstrated how the (multiband) Kondo-lattice model can be used for a
realistic description of the strikingly temperature-dependent electronic excitation
spectrum of local-moment ferromagnets. For this purpose we have combined a many-body
evaluation of the KLM with a \textit{first principles} bandstructure calculation
(TB-LMTO) for the ferromagnetic semiconductors EuO and EuS.

In the first step we inspected an exactly solvable, non-trivial limiting case of the
KLM to learn about the fundamental elementary excitations. In addition, this limiting
case provides a weighty testing criterion for unavoidable approximations in the case of the not
exactly solvable general case.

By use of a self-consistent approach we found out the electronic
as well as magnetic properties of the KLM as functions of decisive model parameters
such as band occupation and interband exchange coupling. By inspecting the Curie
temperature we could demonstrate the transition from a weak-coupling RKKY behaviour to
a strong-coupling double exchange scenario. It appears remarkable that both
features are simultaneously involved in our theory.

Using a straightforward multiband version of the KLM with identical theoretical
approaches we have worked out the influence of the localized magnetic 4f moments of the
Eu$^{2+}$ ion on the 5d conduction bands in EuO and EuS, respectively. The normally
empty bands exhibit a distinct and unconventional (\textit{"not
  Stoner-like"}) temperature-dependent exchange splitting induced by
the interband exchange coupling to the 4f levels. As a special detail we observe the
famous red shift effect in EuO as well as EuS.
\ack
Financial support by the \textit{SFB 290} of the \textit{``Deutsche
  Forschungsgemeinschaft''} is gratefully acknowledged.
%%%%%%%%%%%%%%%%%%%%%%%%%%%%%%%%%%%%%%%%%%%%%%%%%%%%%%%%%%%%%%%%%%%%%% 


\begin{thebibliography}{10}
\expandafter\ifx\csname bibnamefont\endcsname\relax
  \def\bibnamefont#1{#1}\fi
\expandafter\ifx\csname bibfnamefont\endcsname\relax
  \def\bibfnamefont#1{#1}\fi
\expandafter\ifx\csname url\endcsname\relax
  \def\url#1{\texttt{#1}}\fi
\expandafter\ifx\csname urlprefix\endcsname\relax\def\urlprefix{URL }\fi
\providecommand{\bibinfo}[2]{#2}
\providecommand{\eprint}[2][]{\url{#2}}

\bibitem{Wachter79}
\bibinfo{author}{\bibfnamefont{P.}~\bibnamefont{Wachter}},
  \emph{\bibinfo{title}{Handbook of the Physics and Chemistry of Rare Earth}}
  (\bibinfo{publisher}{Amsterdam}, \bibinfo{address}{North Holland},
  \bibinfo{year}{1979}), vol.~\bibinfo{volume}{1}, chap.~\bibinfo{chapter}{19}.

\bibitem{DonDoNo98}
\bibinfo{editor}{\bibfnamefont{M.}~\bibnamefont{Donath}},
  \bibinfo{editor}{\bibfnamefont{P.}~\bibnamefont{Dowben}}, \bibnamefont{and}
  \bibinfo{editor}{\bibfnamefont{W.}~\bibnamefont{Nolting}}, eds.,
  \emph{\bibinfo{title}{Magnetism and Electronic Correlations in Local-Moment
  Systems:Rare-Earth Elements and Compounds}} (\bibinfo{publisher}{World
  Scientific}, \bibinfo{address}{Singapore}, \bibinfo{year}{1998}).

\bibitem{OhnoScience}
\bibinfo{author}{\bibfnamefont{H.}~\bibnamefont{Ohno}},
  \bibinfo{journal}{Science} \textbf{\bibinfo{volume}{281}},
  \bibinfo{pages}{951} (\bibinfo{year}{1998}).

\bibitem{MatsukaraOhno}
\bibinfo{author}{\bibfnamefont{F.}~\bibnamefont{Matsukara}},
  \bibinfo{author}{\bibfnamefont{H.}~\bibnamefont{Ohno}},
  \bibinfo{author}{\bibfnamefont{A.}~\bibnamefont{Shen}}, \bibnamefont{and}
  \bibinfo{author}{\bibfnamefont{Y.}~\bibnamefont{Sugawara}},
  \bibinfo{journal}{Phys.~Rev.~B} \textbf{\bibinfo{volume}{57}},
  \bibinfo{pages}{R2037} (\bibinfo{year}{1998}).

\bibitem{Ramirez97}
\bibinfo{author}{\bibfnamefont{A.~P.} \bibnamefont{Ramirez}},
  \bibinfo{journal}{J.~Phys.: Condens.~Matter} \textbf{\bibinfo{volume}{9}},
  \bibinfo{pages}{8171} (\bibinfo{year}{1997}).

\bibitem{schiller01:_kondo}
\bibinfo{author}{\bibfnamefont{R.}~\bibnamefont{Schiller}},
  \bibinfo{author}{\bibfnamefont{W.}~\bibnamefont{M\"uller}}, \bibnamefont{and}
  \bibinfo{author}{\bibfnamefont{W.}~\bibnamefont{Nolting}},
  \bibinfo{journal}{Phys.~Rev.~B} \textbf{\bibinfo{volume}{64}},
  \bibinfo{pages}{134409} (\bibinfo{year}{2001}).

\bibitem{meyer01:_quant_kondo}
\bibinfo{author}{\bibfnamefont{D.}~\bibnamefont{Meyer}},
  \bibinfo{author}{\bibfnamefont{C.}~\bibnamefont{Santos}}, \bibnamefont{and}
  \bibinfo{author}{\bibfnamefont{W.}~\bibnamefont{Nolting}},
  \bibinfo{journal}{J.~Phys.: Condens.~Matter} \textbf{\bibinfo{volume}{13}},
  \bibinfo{pages}{2531} (\bibinfo{year}{2001}).

\bibitem{NoDu85}
\bibinfo{author}{\bibfnamefont{W.}~\bibnamefont{Nolting}} \bibnamefont{and}
  \bibinfo{author}{\bibfnamefont{U.}~\bibnamefont{Dubil}},
  \bibinfo{journal}{phys.~stat.~sol.~(b)} \textbf{\bibinfo{volume}{130}},
  \bibinfo{pages}{561} (\bibinfo{year}{1985}).

\bibitem{NDM85}
\bibinfo{author}{\bibfnamefont{W.}~\bibnamefont{Nolting}},
  \bibinfo{author}{\bibfnamefont{U.}~\bibnamefont{Dubil}}, \bibnamefont{and}
  \bibinfo{author}{\bibfnamefont{M.}~\bibnamefont{Matlak}},
  \bibinfo{journal}{J.~Phys.: Condens.~Matter} \textbf{\bibinfo{volume}{18}},
  \bibinfo{pages}{3687} (\bibinfo{year}{1985}).

\bibitem{ShasMt81}
\bibinfo{author}{\bibfnamefont{B.~S.} \bibnamefont{Shastry}} \bibnamefont{and}
  \bibinfo{author}{\bibfnamefont{D.~C.} \bibnamefont{Mattis}},
  \bibinfo{journal}{Phys.~Rev.~B} \textbf{\bibinfo{volume}{24}},
  \bibinfo{pages}{5340} (\bibinfo{year}{1981}).

\bibitem{NRMJ97}
\bibinfo{author}{\bibfnamefont{W.}~\bibnamefont{Nolting}},
  \bibinfo{author}{\bibfnamefont{S.}~\bibnamefont{Rex}}, \bibnamefont{and}
  \bibinfo{author}{\bibfnamefont{S.}~\bibnamefont{\mbox{Mathi~Jaya}}},
  \bibinfo{journal}{J.~Phys.: Condens.~Matter} \textbf{\bibinfo{volume}{9}},
  \bibinfo{pages}{1301} (\bibinfo{year}{1997}).

\bibitem{nrrm01}
\bibinfo{author}{\bibfnamefont{W.}~\bibnamefont{Nolting}},
  \bibinfo{author}{\bibfnamefont{G.~G.} \bibnamefont{Reddy}},
  \bibinfo{author}{\bibfnamefont{A.}~\bibnamefont{Ramakanth}},
  \bibnamefont{and} \bibinfo{author}{\bibfnamefont{D.}~\bibnamefont{Meyer}},
  \bibinfo{journal}{Phys.~Rev.~B} \textbf{\bibinfo{volume}{64}},
  \bibinfo{pages}{155109} (\bibinfo{year}{2001}).

\bibitem{nrrmk03}
\bibinfo{author}{\bibfnamefont{W.}~\bibnamefont{Nolting}},
  \bibinfo{author}{\bibfnamefont{G.~G.} \bibnamefont{Reddy}},
  \bibinfo{author}{\bibfnamefont{A.}~\bibnamefont{Ramakanth}},
  \bibinfo{author}{\bibfnamefont{D.}~\bibnamefont{Meyer}}, \bibnamefont{and}
  \bibinfo{author}{\bibfnamefont{J.}~\bibnamefont{Kienert}},
  \bibinfo{journal}{Phys.~Rev.~B} \textbf{\bibinfo{volume}{67}}, \bibinfo{pages}{024426}
  (\bibinfo{year}{2003}).

\bibitem{SaNo02}
\bibinfo{author}{\bibfnamefont{C.}~\bibnamefont{Santos}} \bibnamefont{and}
  \bibinfo{author}{\bibfnamefont{W.}~\bibnamefont{Nolting}},
  \bibinfo{journal}{Phys.~Rev.~B} \textbf{\bibinfo{volume}{65}},
  \bibinfo{pages}{144419} (\bibinfo{year}{2002}).

\bibitem{MNO02}
\bibinfo{author}{\bibfnamefont{W.}~\bibnamefont{M{\"u}ller}} \bibnamefont{and}
  \bibinfo{author}{\bibfnamefont{W.}~\bibnamefont{Nolting}},
  \bibinfo{journal}{Phys.~Rev.~B} \textbf{\bibinfo{volume}{66}},
  \bibinfo{pages}{085205} (\bibinfo{year}{2002}).

\bibitem{SNO01}
\bibinfo{author}{\bibfnamefont{R.}~\bibnamefont{Schiller}} \bibnamefont{and}
  \bibinfo{author}{\bibfnamefont{W.}~\bibnamefont{Nolting}},
  \bibinfo{journal}{Solid State Commun.} \textbf{\bibinfo{volume}{118}},
  \bibinfo{pages}{173} (\bibinfo{year}{2001}).

\bibitem{ChaMil01}
\bibinfo{author}{\bibfnamefont{A.}~\bibnamefont{Chattopadhyay}}
  \bibnamefont{and} \bibinfo{author}{\bibfnamefont{A.~J.}
  \bibnamefont{Millis}}, \bibinfo{journal}{Phys.~Rev.~B}
  \textbf{\bibinfo{volume}{64}}, \bibinfo{pages}{024424}
  (\bibinfo{year}{2001}).

\bibitem{hwang98}
\bibinfo{author}{\bibfnamefont{H.~Y.} \bibnamefont{Hwang}},
  \bibinfo{author}{\bibfnamefont{P.}~\bibnamefont{Dai}},
  \bibinfo{author}{\bibfnamefont{S.~W.} \bibnamefont{Cheong}},
  \bibinfo{author}{\bibfnamefont{G.}~\bibnamefont{Aeppli}},
  \bibinfo{author}{\bibfnamefont{D.~A.} \bibnamefont{Tennant}},
  \bibnamefont{and} \bibinfo{author}{\bibfnamefont{H.~A.} \bibnamefont{Mook}},
  \bibinfo{journal}{Phys.~Rev.~Lett.} \textbf{\bibinfo{volume}{80}},
  \bibinfo{pages}{1316} (\bibinfo{year}{1998}).

\bibitem{andersen75}
\bibinfo{author}{\bibfnamefont{O.~K.} \bibnamefont{Andersen}},
  \bibinfo{journal}{Phys.~Rev.~B} \textbf{\bibinfo{volume}{12}},
  \bibinfo{pages}{3060} (\bibinfo{year}{1975}).

\bibitem{andersen84}
\bibinfo{author}{\bibfnamefont{O.~K.} \bibnamefont{Andersen}} \bibnamefont{and}
  \bibinfo{author}{\bibfnamefont{O.}~\bibnamefont{Jepsen}},
  \bibinfo{journal}{Phys.~Rev.~Lett.} \textbf{\bibinfo{volume}{53}},
  \bibinfo{pages}{2571} (\bibinfo{year}{1984}).

\bibitem{schiller01:_predic_euo}
\bibinfo{author}{\bibfnamefont{R.}~\bibnamefont{Schiller}} \bibnamefont{and}
  \bibinfo{author}{\bibfnamefont{W.}~\bibnamefont{Nolting}},
  \bibinfo{journal}{Phys.~Rev.~Lett.} \textbf{\bibinfo{volume}{86}},
  \bibinfo{pages}{3847} (\bibinfo{year}{2001}).

\bibitem{gelfert01}
\bibinfo{author}{\bibfnamefont{A.}~\bibnamefont{Gelfert}} \bibnamefont{and}
  \bibinfo{author}{\bibfnamefont{W.}~\bibnamefont{Nolting}},
  \bibinfo{journal}{J.~Phys.: Condens.~Matter} \textbf{\bibinfo{volume}{13}},
  \bibinfo{pages}{R505} (\bibinfo{year}{2001}).

\bibitem{batlogg75}
\bibinfo{author}{\bibfnamefont{B.}~\bibnamefont{Batlogg}},
  \bibinfo{author}{\bibfnamefont{E.}~\bibnamefont{Kaldis}},
  \bibinfo{author}{\bibfnamefont{A.}~\bibnamefont{Schlegel}}, \bibnamefont{and}
  \bibinfo{author}{\bibfnamefont{P.}~\bibnamefont{Wachter}},
  \bibinfo{journal}{Phys.~Rev.~B} \textbf{\bibinfo{volume}{12}},
  \bibinfo{pages}{3940} (\bibinfo{year}{1975}).

\bibitem{BDK80}
\bibinfo{author}{\bibfnamefont{H.}~\bibnamefont{G.}~\bibnamefont{Bohn}},
  \bibinfo{author}{\bibfnamefont{W.}~\bibnamefont{Zinn}},
  \bibinfo{author}{\bibfnamefont{B.}~\bibnamefont{Dorner}}, \bibnamefont{and}
  \bibinfo{author}{\bibfnamefont{A.}~\bibnamefont{Kollmar}},
  \bibinfo{journal}{Phys.~Rev.~B} \textbf{\bibinfo{volume}{22}},
  \bibinfo{pages}{5447} (\bibinfo{year}{1980}).
\end{thebibliography}
\end{document}